\documentclass[aps,pra,showpacs,groupedaddress,superscriptaddress,twocolumn]{revtex4}
\usepackage{amsfonts}
\usepackage{graphicx}
\usepackage{latexsym}   
\usepackage{makeidx} 
\usepackage{amsmath}
\usepackage{amssymb}  
\usepackage{graphicx}
\usepackage{color} 
\begin{document}

\title{First-order superfluid-Mott-insulator transition for quantum optical switching \\ in cavity QED arrays with two cavity modes}
\author{Kenji Kamide}
\affiliation{%
Department of Physics, Osaka University, Toyonaka, Osaka 560-0043, Japan
}
\email{kamide@acty.phys.sci.osaka-u.ac.jp}
\author{Makoto Yamaguchi}
\affiliation{%
Department of Physics, Osaka University, Toyonaka, Osaka 560-0043, Japan
}
\author{Takashi Kimura}
\affiliation{%
Department of Mathematics and Physics, Kanagawa University, 2946 Tsuchiya, Hiratsuka, Kanagawa 259-1293, Japan}
\author{Tetsuo Ogawa}%
%\homepage{http://wwwacty.phys.sci.osaka-u.ac.jp/~ogawa/cover.html}
\affiliation{%
Department of Physics, Osaka University, Toyonaka, Osaka 560-0043, Japan
}%
%\affiliation{
\date{\today}% It is always \today, today,
\begin{abstract}
We theoretically investigated the ground states of coupled arrays of cavity quantum electrodynamical (cavity QED) systems in presence of two photon modes. Within the Gutzwiller-type variational approach, we found the first-order quantum phase transition between Mott insulating and superfluid phases as well as the conventional second-order one.
The first-order phase transition was found only for specific types of emitter models, and its physical origin is clarified based on the analytic arguments which are allowed in the perturbative and semiclassical limits. 
The first-order transition of the correlated photons is accompanied with discontinuous change in the emitter states, not only with the appearance of inter-cavity coherence in the superfluid phase.
We also discuss the condition for the first-order transition to occur, which can lead to a strategy for future design of quantum optical switching devices with cavity QED arrays.
\end{abstract}
\maketitle 

\section{\label{sec1}Introduction}
Recently, parametric controls of quantum optics through manipulating states of matter in quantum optical systems have become key technologies to develop a new quantum optical device.
In the studies, a cavity quantum electrodynamical (cavity QED) system, where photons interact with one or $N$ emitters in one cavity, has been known to exhibit striking physics such as the strong nonlinearity, open-dissipative nature, and phase transitions~\cite{Kimble,Fushman,Valle,Ota,Hennessy, YamaguchiR, Strauf, Nomura, Dicke, Hepp, Emary}.
%In the studies, a cavity quantum electrodynamical (cavity QED) system, where photons interact with one or $N$ emitters in one cavity, has been known to exhibit striking quantum optical features: the strong nonlinearity e.g. the photon blockade~\cite{Kimble} and strong photon-photon interaction~\cite{Fushman}, two-photon processes~\cite{Valle,Ota}, open dissipative effects of the quantum optics~\cite{Hennessy, YamaguchiR}, and various types of phase transitions such as lasing from a single emitter (quantum dot)~\cite{Strauf, Nomura} and superradiant phase transition~\cite{Dicke, Hepp, Emary}.
In this context, coupled cavity QED arrays have also been focused so much in recent years, due to the new possibility toward a manipulation of the quantum optics with an efficient use of the many-body feature of photons after pioneering papers appeared~\cite{Greentree,Hartmann, Angelakis, HartmannR}. 
Toward the realization, the effective coupling between cavity arrays was recently obtained with photonic crystal microcavities~\cite{Rundquist}, although its strong light-matter coupling regime has not been reached yet.
The simplest case of the cavity QED arrays is described by Jaynes-Cummings Hubbard (JCH) model, where the photons can hop to the neighboring cavities while they suffers repulsive interaction with each other through the light-matter interaction.

Coupled cavity QED arrays have been shown to exhibit the superfluid (SF)-Mott-insulator (MI) transition of photons~\cite{Greentree,Hartmann,Angelakis}, being similar to Bose-Hubbard (BH) model which has long been studied for Josephson junction arrays and neutral Bose gases in optical lattices~\cite{Fisher,Batrouni,Elstner,Polak,Teichmann,Stoof, Greiner}.  
Therefore, they have been considered as a new candidate of quantum simulator of many-body physics in solids.
As well as the scientific interests, the quantum phase transition of photons is important also on its application purpose for a new source of quantum-correlated photons.

%Recently, numbers of papers in this issue focus on the other aspects on the pure physics; accurate treatment of the many-body correlation and fluctuation spectra~\cite{Rossini,Schmidt, Aichhorn, Pippan}, the absence of Mott lobes in ultra-strong coupling regime~\cite{Takada,Treci}, the non-equilibrium feature induced by pump and loss~\cite{Nissen,Knap}, and the nonequilibrium phase transition~\cite{Tomadin,FazioR}.
%Most of them are aimed for scientific interests where the cavity QED arrays has been considered as a new candidate of quantum simulator of many-body physics in solids.
Recently, numbers of papers in this issue focus on further aspects on the pure physics e.g. the quantum fluctuations, ultrastrong coupling physics, and nonequilibrium physics~\cite{Rossini,Schmidt, Aichhorn, Pippan,Takada,Treci,Nissen,Knap,Tomadin,FazioR}.
On the other hand, the technological applications of cavity QED arrays have not been focused so much.
For applications, the quantum phase transition of the coupled cavity QED arrays seems to be useful for optical switching~\cite{Englund} and sensing devices with high sensitivity.
However, in order to obtain a high performance of the switching, the phase transition of the first order will be more feasible than the conventional second-order one for the simple JCH model.
Therefore, important questions to clarify for applications are now, (I) does the first-order transitions exist in cavity QED arrays, and if yes, (II) what is the condition for the first-order transitions to arise?
In this regard, it is known in BH models (with a close relation to cavity QED arrays~\cite{JKoch}) that the SF-MI transition can be of the first order if the multiple components of bosons are present~\cite{Kimura,KimuraJPS,Krutitsky,Kuklov}. It was also shown in spin-1 bosons that the first-order transition is only found for antiferromagnetic spin-spin interaction~\cite{Kimura}. Therefore, both the multiple components of bosons and the types of boson-boson interactions will be the key ingredients for the first-order transition to occur.

\begin{figure}[!tbp] 
\begin{center}
\includegraphics[width=.4\textwidth]{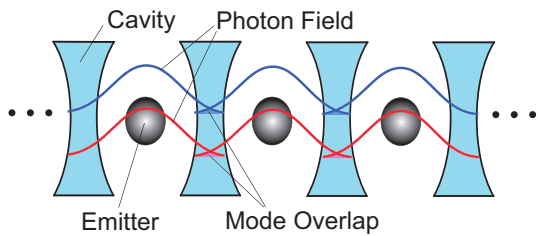} {}
\end{center}
\caption{\label{fig1} (Color online) Coupled cavity QED array with two cavity modes.} 
\end{figure}

In this paper, we study the SF-MI transition in a variety of coupled cavity QED arrays in presence of two cavity modes (Fig.~\ref{fig1}), as a simplest case of multiple component bosons. We find the first-order SF-MI transitions can occur also in cavity QED arrays and can be applicable to optical switching devices.
The Hamiltonian used here is generally given by
\begin{eqnarray}
\hat{H}=\sum_{i={\rm site}} \hat{h}_{0,i}-t\sum_{i,j}\sum_{m=A,B}
 \left( \hat{a}_{m,i}^\dagger \hat{a}_{m,j} + {\rm h.c.} \right). \label{eq:GenModel}
\end{eqnarray}
where the first and second terms are the Hamiltonian within each cavity $i$ and hopping of photons of modes $m$ ($=A, B$) between neighboring cavities $i$ and $j$.
The types of the light-matter interactions are varied with a choice of $\hat{h}_{0,i}$.
By applying Gutzwiller-type variational approach~\cite{Vollhardt} for the ground states of coupled cavity QED arrays, we show a certain range of the models exhibit a first-order phase transition.
Therefore, we conclude that the types of light-matter interactions are important for the first-order transition to occur also in this coupled QED arrays, being consistent with the case of spin-1 BH models~\cite{Kimura,KimuraJPS,Krutitsky}.

This paper is organized as follows. 
In Sec.~\ref{sec2}, we show detailed results of the Gutzwiller-type variational calculation for a $\Lambda$-type three-level configuration of emitters as an example of the model in Eq.~(\ref{eq:GenModel}), where the first-order transition is found.
In Sec.~\ref{sec3}, we show results for four different types of emitter models where the first-order transition is found only for specific types. 
Conditions for the first-order transition to occur are discussed. 
%In Sec.~\ref{sec4}, we briefly show results for a two-component Bose-Hubbard model and shows that the first-order transition can be found only for antiferromagnetic interaction between pseudo spins being similar to spin 1 case~\cite{Ref4}. By mapping the QED array systems to effective two-component Bose-Hubbard models and determining the effective interaction parameters, we will discuss the condition for the first-order transition to be found.
Finally, we will summarize the results, conclusions, and future remarks in Sec.~\ref{sec5}.

\section{\label{sec2}Two-mode cavity QED arrays with $\Lambda$-type three level emitters ($\Lambda$1)} 
In this section, we focus on the quantum phase transitions in two-mode cavity QED arrays with $\Lambda$-type emitters, as a simple example showing the first-order phase transitions. 
\subsection{Dressed states of $\Lambda$-type system ($t = 0$)}
\begin{figure}[!tbp]
\begin{center}
\includegraphics[width=.35\textwidth]{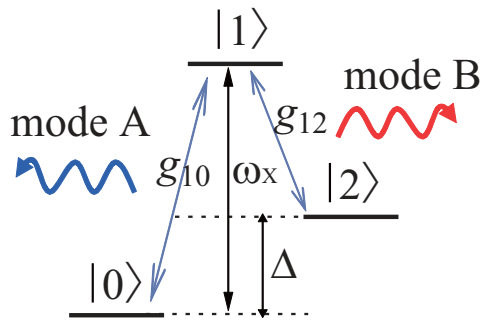} {}
\end{center}
\caption{\label{fig2}(Color online) Energy diagram for $\Lambda$-type three level emitters.}
\end{figure}

We will first show the detailed results for a two-mode cavity QED array system with $\Lambda$-type three level emitters whose energy level diagram inside a cavity is shown in Fig.~\ref{fig2}. We call here this model ``$\Lambda$1''.
For this configuration, photons of a mode A couples with a level transition between $|0 \rangle$ and $|1 \rangle$, and a mode B couples with a level transition between $|1 \rangle$ and $|2 \rangle$. The Hamiltonian $\hat{h}_{0,i}$ in Eq.~(\ref{eq:GenModel}) is given by
\begin{eqnarray}
\hat{h}_0&=&\omega_X|1\rangle \langle  1 |+\Delta|2\rangle \langle  2 |+\omega_A \hat{a}_A^\dagger \hat{a}_A +\omega_B \hat{a}_B^\dagger \hat{a}_B   \nonumber \\
&+& \left( g_{10}|1\rangle \langle  0 | \hat{a}_A+g_{12}|1 \rangle \langle 2 | \hat{a}_B 
+{\rm h.c.} \right) -\mu \hat{\mathcal{N}}_{tot}, \label{eq:HLambda}
\end{eqnarray}
where $\hat{a}_A$ and $\hat{a}_B$ represent annihilation operators of cavity modes A and B. The site index $i$ is omitted for simplicity. 
The total excitation number per cavity $\hat{\mathcal{N}}_{tot} \equiv |1\rangle \langle  1 | +\hat{a}_A^\dagger \hat{a}_A+\hat{a}_B^\dagger \hat{a}_B$ is conserved with this Hamiltonian $\hat{h}_0$ and the mean excitation number is fixed by a chemical potential $\mu$.
Throughout this paper, we assume that $\mu$ measures the strength of the energy injection by a external pump bath.
Here we also assume the light-matter coupling constants $g_{10}=g_{12}\equiv g$ and resonance conditions for cavity modes, $\omega_A=\omega_X$ and $\omega_B=\omega_X-\Delta$, for simplitity.
We note that this simplification does not change the discussions.
Without loss of generality, we set $\Delta>0$ in this paper.

\begin{figure}[!tbp]
\begin{center}
\includegraphics[width=0.35\textwidth]{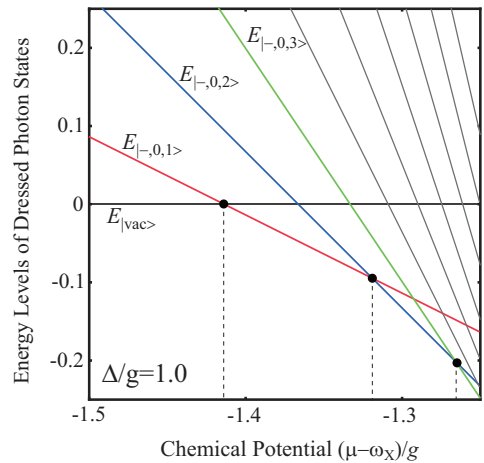} {}
\end{center}
\caption{\label{fig3}(Color online) Lowest energy levels i.e. the eigenvalues of Eq.~(\ref{eq:HLambda}), for $\Lambda$-type three level emitters for $\Delta/g=1.0$.}
\end{figure}

Without tunneling $t=0$ in Eq.~(\ref{eq:GenModel}), the energy levels of this $\Lambda$-type cavity QED system are given by the eigenvalues of Eq.~(\ref{eq:HLambda}) and shown in Fig.~\ref{fig3}.
In the following, we use a vector $|\alpha, n_A,n_B \rangle$ with an emitter state $\alpha$ ($=1,2,3$) and numbers of photon modes $n_A$ and $n_B$ (eigenvalues of ${\hat n}_A$ $(\equiv \hat{a}_A^\dagger \hat{a}_A)$ and ${\hat n}_B$ $(\equiv \hat{a}_B^\dagger \hat{a}_B)$ equal to $0,1,\cdots$) that forms the complete basis set of this Hamiltonian.
The eigenstates of the Hamiltonian $\hat{h}_0$ consist of dressed states and decoupled states.
There are three types of dressed states given by
\begin{eqnarray}
|{\rm M},n_A,n_B \rangle &=&  \sqrt{\frac{1+n_B}{2+n_A +n_B}}|0,n_A+1,n_B \rangle \nonumber \\
&& - \sqrt{\frac{1+n_A}{2+n_A +n_B}}|2,n_A,n_B+1 \rangle ,
\end{eqnarray}
and
\begin{eqnarray}
&& |\pm,n_A,n_B \rangle \nonumber \\
&=& \pm \sqrt{\frac{1}{2}}|1,n_A,n_B \rangle
+ \sqrt{\frac{1+n_A}{4+2 n_A +2 n_B}}  |0,n_A+1,n_B \rangle \nonumber \\
&& + \sqrt{\frac{1+n_B}{4+2 n_A +2 n_B}} |2,n_A,n_B+1 \rangle, \label{eq:DressdPhoton1}
\end{eqnarray}
with the eigen energies
\begin{eqnarray}
E_{|{\rm M}, n_A,n_B \rangle}&=&(\omega_X-\mu)(n_A+n_B+1)-\Delta n_B, \\
E_{|\pm, n_A,n_B \rangle}&=&(\omega_X-\mu)(n_A+n_B+1)-\Delta n_B \nonumber \\
&& \pm g \sqrt{n_A+n_B+2}.
\end{eqnarray}
The remaining decoupled states are 
\begin{eqnarray}
|0, 0, n_B \rangle \ {\rm and} \  |2, n_A 0 \rangle,
\end{eqnarray}
with the eigen energies
\begin{eqnarray}
E_{|0, 0, n_B \rangle}&=&(\omega_X-\mu-\Delta)n_B, \\
E_{|2, n_A,0 \rangle}&=&(\omega_X-\mu)n_A + \Delta,
\end{eqnarray}
Figure \ref{fig3} shows that the ground state changes from the vacuum state $|{\rm vac} \rangle \equiv |0,0,0 \rangle$ to dressed photon state  $|-,0,n_B \rangle$ ($n_B=0,1, \cdots$) with larger total excitation number $\mathcal{N}_{tot}=n_B+1$ ($\mathcal{N}_{tot}$ is the eigenvalue of $\hat{\mathcal{N}}_{tot}$) as the chemical potential $\mu$ is increased. For $\omega_X -\mu -\Delta>0$, the ground state changes at level crossing points $\mu=\mu_{c}$ with $(\mu_{c}-\omega_X)/g=-\sqrt{2}, -\Delta/g-(\sqrt{3}-\sqrt{2})$, $-\Delta/g-(\sqrt{4}-\sqrt{3})$, $\cdots$ ($\approx -1.414, -1.318, -1.268, \cdots$ for $\Delta/g=1.0$).

\subsection{\label{sec2-2} Ground state phase diagram ($t \ne 0$)}

Now we will find the ground state of the cavity QED arrays in case of $t \ne 0$ when the hopping of cavity photons is effective.
Using a Gutzwiller-type variational wavefunctions for the coupled cavity system
\begin{eqnarray}
| \Phi \rangle&=&\prod_{i={\rm site}}| \Phi_i \rangle, \\
| \Phi_i \rangle&=& \sum_{\alpha,n_A,n_B} g(\alpha, n_A, n_B) | \alpha, n_A,n_B \rangle, 
\end{eqnarray}
where the ground states of a product state with site independent $| \Phi_i \rangle$.
We minimize the expectation value of the free energy $\langle \Phi |\hat{H} | \Phi \rangle$ to find the variational parameter $g(\alpha, n_A, n_B)$. 
This approach gives the same result as the mean field approach~\cite{Greentree} which is to approximate Eq.~(\ref{eq:GenModel}) by a mean field Hamiltonian 
\begin{eqnarray}
&&\hat{H}^{MF}(\psi_A, \psi_B)=\sum_i \hat{h}_{0,i}  \nonumber \\
&&\quad -zt\sum_{i,m}\left(\psi_m^\ast \hat{a}_{m,i}+\psi_m \hat{a}_{m,i}^\dagger -|\psi_m|^2 \right), \label{eq:MFHamiltonian}
\end{eqnarray}
and minimizing its lowest eigen energy with respect to site-independent variational parameters $\psi_A$ and $\psi_B$ defined by $\psi_A \equiv \langle \hat{a}_{A,i} \rangle$, $\psi_B \equiv \langle \hat{a}_{B,i} \rangle$. In Eq.~(\ref{eq:MFHamiltonian}), $z$ denotes a number of cavities of nearest neighbors.

\begin{figure}[!tbp]
\begin{center}
\includegraphics[width=0.48\textwidth]{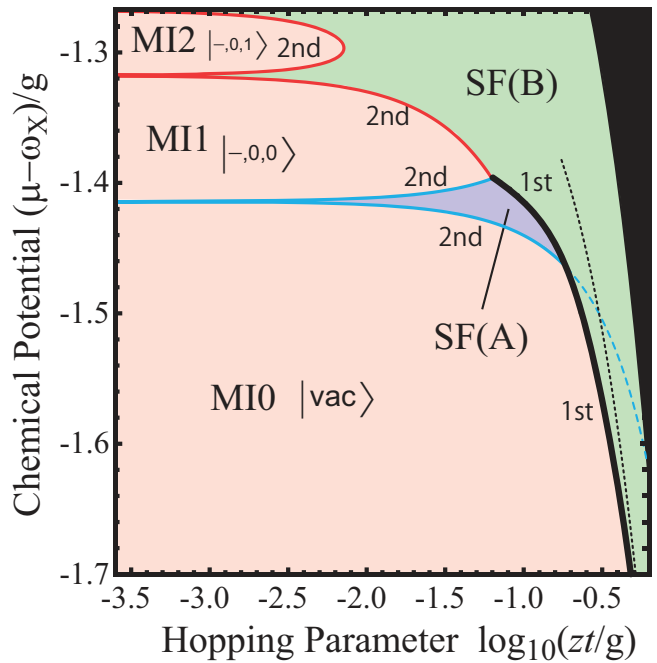} {}
\end{center}
\caption{\label{fig4} (Color online) Ground state phase diagram of two-mode coupled cavity QED arrays with $\Lambda$-type emitters obtained by the Gutzwiller-type variational wavefunction approach for $\Delta/g=1.0$. The order of the phase transition are indicated at the phase boundaries, (bold, black) and (solid, blue/red), by ``1st'' and ``2nd'', respectively. All the second-order phase boundary is given by a perturbation theory in Eq.~(\ref{eq:PBsecond}), whereas the dashed line (blue) inside SF(B) phase is a part of the curve given by Eq.~(\ref{eq:PBsecond}) plotted for a guide. Dotted line (black) in SF(B) phase shows a prediction of the first-order phase boundary between MI0 and SF(B) phases within the semiclassical approximation in Eq.~(\ref{eq:SemiPB_lambda}). The dotted line is shown only for the low-$\mu$ region near the MI0 phase boundary where Eq.~(\ref{eq:SemiPB_lambda}) is meaningful.}
\end{figure} 
\begin{figure}[!tbp] 
\begin{center}
\includegraphics[width=0.5\textwidth]{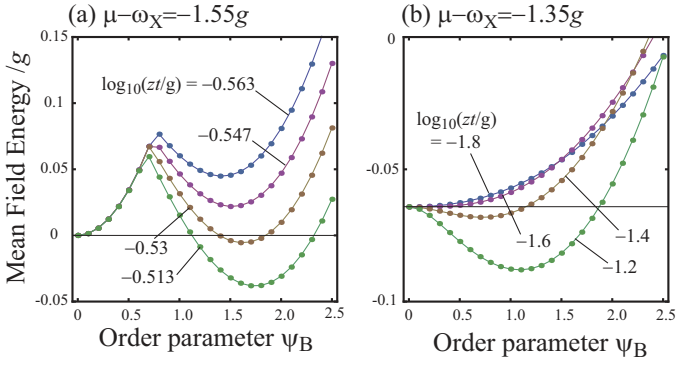} {}
\end{center}
\caption{\label{fig5} (Color online) Mean field energies are plotted as a function of a variational order parameter $\psi_B=\langle a_B \rangle$ for (a) $(\mu-\omega_X)/g=-1.55$ and (a) $(\mu-\omega_X)/g=-1.35$. We set $\psi_A=0$ in this figure.}
\end{figure}

In Fig.~\ref{fig4}, the ground state phase diagram is obtained numerically with a cutoff $n_{\rm max}$ ($=15$) to the photon number basis so that $0 \le n_A, n_B \le n_{\rm max}$. In the black area determined by $\omega_X-\mu-\Delta-zt<0$, the number of photons diverges for $n_{\rm max} \to + \infty$ and we can no longer determine a stationary state with a fixed mean-number~\cite{JKoch}.
The phase diagram contains three phases: MI phases with $\psi_A=\psi_B= 0 $ and quantized number states of $\mathcal{N}_{tot}$, and two SF phases, SF(A) with $\psi_A \ne 0 $ and $\psi_B =0$, and SF(B) with $\psi_A = 0 $ and $\psi_B \ne 0$.
Since the effective mean field Hamiltonian is equal to Eq.~(\ref{eq:GenModel}) with $t=0$ in MI phases, the ground states are the vacuum states $|\rm vac \rangle$ with $\mathcal{N}_{tot}=0$ or dressed photon states $|-,0,n_B \rangle$ with $\mathcal{N}_{tot}=n_B+1$.
As indicated at the phase boundaries in  Fig.~\ref{fig4}, the phase transition can be of the second or first order.

For high $\mu$ (high excitation density), the phase transition is always the second-order one between MI with $\mathcal{N}_{\rm tot}\ge 1$ and SF phases (A or B). 
The second-order phase boundary curves between MI and SF phases are also given by a second-order perturbation theory with respect to the small $zt\psi$'s in Eq.~(\ref{eq:MFHamiltonian}). The energy correction per cavity ($\delta E_{l}$) to the unperturbed MI states ($l={\rm MI0}, \ {\rm MI1}, \ {\rm MI2} \cdots $) has the following form:
\begin{eqnarray}
&&\delta E_{l}(\psi_A,\psi_B)=\left(zt-z^2 t^2 C_{A,l} \right) |\psi_A|^2 \nonumber \\
&& \qquad +\left(zt-z^2 t^2 C_{B,l} \right) |\psi_B|^2 +\mathcal{O}\left((tz\psi)^4 \right), \label{eq:Emfsecond}
\end{eqnarray}
where the analytic expression of the coefficients $C_{A,l}$ and $C_{B,l}$ are given in Table \ref{table1} in Appendix \ref{Sec:Perturbation}. 

If $C_{A,l}>C_{B,l}$, there are three possible cases: (i) $zt<1/C_{A,l}$ where the coefficients of $|\psi_A|^2$ and $|\psi_B|^2$ in Eq.~(\ref{eq:Emfsecond}) are all positive, (ii) $1/C_{A,l}<zt<1/C_{B,l}$ where the coefficients are negative for $|\psi_A|^2$ and positive for $|\psi_B|^2$, and (iii) $1/C_{B,l}<zt$ where the coefficients of $|\psi_A|^2$ and $|\psi_B|^2$ are both negative. This indicates that the ground state is the MI state ($\psi_A=\psi_B=0$) for (i), and SF(A) state ($\psi_A > 0$ and $\psi_B=0$) for (ii). As for (iii), the perturbative expression in Eq.~(\ref{eq:Emfsecond}) around MI state does not give any reliable prediction. Therefore, the second-order phase transition from MI to SF(A) occurs at $zt=1/C_{A,l}$ for $C_{A,l}>C_{B,l}$. 
We also find, by applying a similar argument for $C_{B,l}>C_{A,l}$, that the second-order phase transition from MI to SF(B) occurs at $zt=1/C_{B,l}$ for $C_{A,l}>C_{B,l}$.
As a result, the phase boundary of the second-order transition from MI to SF(A)/SF(B) is given by
\begin{eqnarray}
zt={\rm min}\left(1/C_{A,l}, 1/C_{B,l} \right). \label{eq:PBsecond}
\end{eqnarray}

On the other hand, for low $\mu$, the phase transition between the MI phase with $\mathcal{N}_{\rm tot}=0$ and SF(B) phase, and that between SF(A) and SF(B) phases are first-order transitions in Fig.~\ref{fig4}.
The first-order transition occurs with a discontinuous jump in the variational parameters i.e. a discontinuous change in the ground state.
Therefore we have to deal with a nonzero value of $\psi$ in order to determine the phase boundary which cannot be obtained by the second-order perturbation theory from the MI phases. 
Instead, we evaluated numerically the mean-field energy profile in ($\psi_A , \psi_B$)-plane which shows the distinct two energy minima in Fig.~\ref{fig5} (a), and the position of the global minimum changes from $\psi_B=0$ to $\psi_B \approx 1.6$ at a critical hopping parameter $t=t_c$ ($-0.547<\log_{10}(zt_c/g)<-0.53$). 
If the same applies to the case of the second-order transition (Fig.~\ref{fig5} (b)), the energy profile shows only one minimum that smoothly changes with hopping parameter.

In the last of this subsection, we will refer to the dimensionality of the system. 
Within the mean-field approach presented here, the dimensionality of the cavity arrays is incorporated only through the coordination number~$z$.
Therefore, our mean-field results can be used for any dimensional arrays. 
However, it is known that the mean-field approach overestimates the superfluid phases compared with exact numerical methods including quantum fluctuations such as quantum Monte Carlo method. 
The deviation of mean-field critical values $zt_c/g$ from the more accurate results can be estimated as that of 10 percent in two or three dimensional systems~\cite{Schmidt}, while it can be more than 200 percent in one dimensional arrays due to enhanced quantum fluctuations~\cite{Rossini}. Therefore, we consider our conclusions are applicable to two and three dimensional cavity arrays.

\subsection{Classical and Quantum optics, and Photoluminescence spectra}

\begin{figure}[!tbp] 
\begin{center}
\includegraphics[width=.50\textwidth]{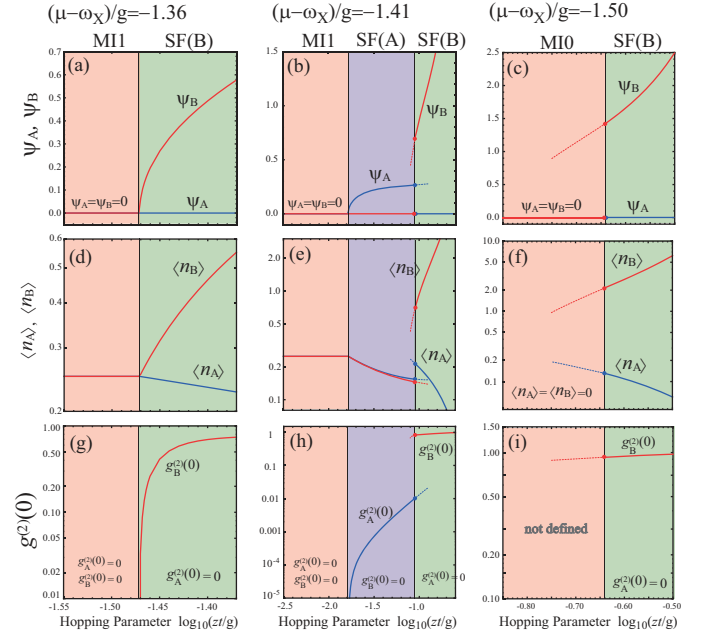} {}
\end{center}
\caption{\label{fig6} (Color online) Amplitudes, $ \psi_A$ and $\psi_B$, mean numbers, $ \langle \hat{n}_A \rangle $ and $\langle \hat{n}_B \rangle$, and second-order correlation at zero time delay, $g_A^{(2)}(\tau=0)$ and $g_B^{(2)}(\tau=0)$, are evaluated as a function of hopping parameters. The dotted curves extended over the first-order phase boundary show the values for the metastable states (local minima of the mean field energy), and the end of the dotted curves indicate where the metastable states have disappeared.} 
\end{figure}

In Fig.~\ref{fig6}, we show the amplitude $\psi_{m}$ [(a-c)], mean numbers $\langle \hat{n}_m \rangle $ [(d-f)], and second-order number correlation at zero time delay $g^{(2)}_{m}(0)=\langle \hat{a}_m^\dagger \hat{a}_m^\dagger \hat{a}_m \hat{a}_m \rangle/\langle \hat{n}_m \rangle^2$ (which is not defined for a vacuum state) [(g-i)] for cavity modes $m=A, B$.  They are plotted as a function of hopping parameter for different chemical potentials: $(\mu-\omega_X)/g=-1.36$ (left row), $-1.41$ (middle row), and $-1.50$ (right row).
Depending on the chemical potential, we found different types of phase transitions: A second-order transition from MI1 to SF(B) for $(\mu-\omega_X)/g=-1.36$, a second-order transition from MI1 to SF(A) and a first-order transition from SF(A) to SF(B) for $(\mu-\omega_X)/g=-1.41$, and a first-order transition from MI0 to SF(B) for $(\mu-\omega_X)/g=-1.50$.
The signatures of the first-order transition are the discontinuous jumps in these figures, which can be applicable to switching devices of the classical and quantum optics.
For example, the luminescence from mode B will show a strong antibunching ($g^{(2)}_{B}(0)=0$) at $\log_{10}(zt/g)=-1.05-0$ and almost perfect coherence ($g^{(2)}_{B}(0) \approx 1$) at $\log_{10}(zt/g)=-1.05 +0$ in Fig.~\ref{fig6} (h).
As being the typical first-order transition, metastable states are found near the first-order phase boundary. In Fig.~\ref{fig6} (b), (e), (h), (c), (f), and (i), the dotted curves extended over the first-order phase boundary show the plots for the metastable states i.e. the local minima of the mean field energy, and the end of the dotted curves indicate where the metastable states have disappeared and the local minima turned into local maxima.

\begin{figure}[!tbp] 
\begin{center}
\includegraphics[width=.48\textwidth]{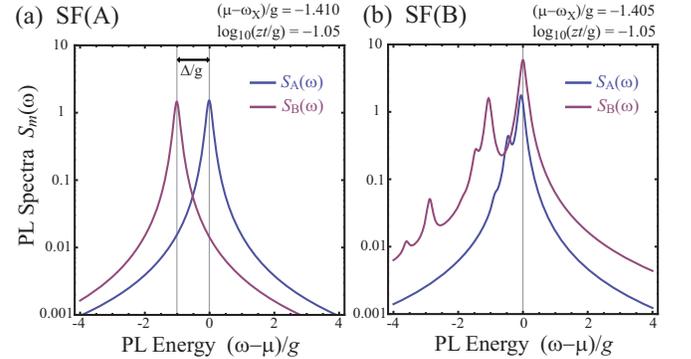} {}
\vspace{-10mm}
\end{center}
\caption{\label{fig7} (Color online) Photoluminescence spectra of $\Lambda$1 model ($\Delta/g=1.0$), $S_A(\omega)$ and $S_B(\omega)$ in Eq.~(\ref{eq:PL_Lambda1}), are shown by blue and red lines respectively, for different parameters in SF(A) and SF(B) of Fig.~\ref{fig4}, which are located on either side of the first-order phase boundary at $\log_{10}(zt/g)=-1.05$: (a) SF(A) phase for $(\mu-\omega_X)/g=-1.410$, and (b) SF(B) phase for $(\mu-\omega_X)/g=-1.405$.
For the plots, the detector resolution is $\gamma_r/g=0.1$. } 
\end{figure}

The calculated photoluminescence (PL) spectra are shown in Fig.~\ref{fig7}.
Two figures, (a) and (b), are obtained for the parameters $(\mu-\omega_X)/g=-1.4075 \pm \delta$ with $\delta=0.0025$ on either side of the first-order phase boundary between SF(A) and SF(B) phases.
With such small change in parameters, the large difference is obtained, and we see the first-order transition can be used also for a classical optical switching.
The PL spectra are obtained using the linear response theory by assuming that the photon of mode $m$ ($=A,B$) leaks out of the QED arrays very weakly~\cite{Scully}, and given by
\begin{eqnarray}
S_m(\omega)&\propto&{\rm Re}\int_0^\infty \  \langle G|  \hat{a}^\dagger_m (\tau) \hat{a}_m |G \rangle e^{-i\omega \tau -\gamma_r \tau}{\rm d}\tau \nonumber \\
&=& \pi \sum_l \frac{\gamma_r |\langle l | \hat{a}_m |G \rangle|^2}{(\omega+E_l-E_G)^2+ (\gamma_r)^2}, \label{eq:PL_Lambda1}
\end{eqnarray}
where $|G \rangle$ ($|l \rangle$) and $E_{G}$ ($E_{l}$) are the eigen state and eigen energy of the ground (excited) state for the mean field Hamiltonian in Eq.~(\ref{eq:MFHamiltonian}) given by the solution of the variational problem of $\psi$'s. The Heisenberg representation is used for $\hat{a}^\dagger_m (\tau)=\exp[i\hat{H}^{MF}\tau] \hat{a}^\dagger_m \exp[-i\hat{H}^{MF}\tau]$.
A phenomenological parameter, $\gamma_r$, is introduced here to account for a resolution of the detector.  
In Fig.~\ref{fig7} (a) obtained in the SF(A) phase, the PL from the modes A and B shows double Lorentzian peaks. They can be understood as spontaneous emission.
This is consistent with the observation that $g^{(2)}_A(0)$ and $g^{(2)}_B(0)$ are much less than 1 in SF(A) phase in Fig.~\ref{fig6} (h).
On the other hand, in Fig.~\ref{fig7} (b) obtained in SF(B) phase, a strong main peak and the subpeak structure are newly found at the chemical potential ($\omega=\mu$) and the lower energy side ($\omega<\mu$), respectively. 
The strong main peak corresponds to the coherent emission from the Bose condensate of mode B, and the subpeak structure correspond to the Mollow-like side peaks.
In the Mollow triplet in quantum optics~\cite{Mollow}, which are found for resonant fluorescence from an emitter driven by classical field in an out-of-equilibrium condition, the side peaks are found in both the higher and lower side of the main peak. On the other hand, in our case, the whole system of photons and emitters are in a ground state of the thermal equilibrium. Due to the difference, the subpeak structure should be found only in the lower side of the main peak (See Appendix \ref{Sec:Mollow}).

\subsection{Origin of the first-order phase transition}

Here we discuss the physics why the first-order transition appears in this system (especially for small $\mu$ region).

In this two-mode cavity system coupling with $\Lambda$-type emitters, we find only the two superfluid phase SF(A) and SF(B), while we find no SF(A+B) phase with $\psi_A \ne 0$ and $\psi_B \ne 0$ (which can arise in different emitter models as shown in the next section).
We can naively understand why there is no SF(A+B) phase in this system by the following semiclassical analysis (which can be reliable in presence of sufficiently large amplitude of photons $|\psi_m|\gg 1$ inside SF phases) in combination with the complementary results based on the perturbation theory given in Sec.~\ref{sec2-2}.
In the semiclassical approximation (putting $\psi_m$ into $a_m$ in Eq.~(\ref{eq:GenModel}) and Eq.~(\ref{eq:HLambda})), the lowest eigen energy $E_{\rm semi}$ of the system is 
\begin{eqnarray} 
E_{\rm semi} &=& (\omega_X -\mu -zt)|\psi_A|^2 +(\omega_X -\mu -zt-\Delta)|\psi_B|^2 \nonumber \\
&&    -g\sqrt{|\psi_A|^2+|\psi_B|^2} 
+E_{\rm emit} +\mathcal{O}(1/\psi ),  \ \label{eq:Esemi_lambda} 
\end{eqnarray}
where the last term $E_{\rm emit}$ is a small contribution from the average of total energy of the emitters:
\begin{eqnarray} 
H_{\rm emit} \equiv   (\omega_X -\mu)\hat{P}_1+\Delta \hat{P}_2,
\end{eqnarray}
with $\hat{P}_{\alpha}\equiv |\alpha \rangle \langle \alpha|$ being a projection to the emitter state $\alpha=0,1,2$. 
In the semiclassical limit $|\psi_A|^2+|\psi_B|^2 \gg 1$, it is explicitly given by
\begin{eqnarray}
E_{\rm emit} = \frac{\omega_X-\mu}{2}+\frac{\Delta}{2} \frac{|\psi_B|^2}{|\psi_A|^2+|\psi_B|^2},
\end{eqnarray}
which is small compared to other terms in Eq.~(\ref{eq:Esemi_lambda}). Neglecting the small contribution, the saddle point of $E_{\rm semi}$ is determined by
\begin{eqnarray}
(\omega_X -\mu -zt)\psi_A&=&\frac{g\psi_A/2}{\sqrt{|\psi_A|^2+|\psi_B|^2}} ,\\
(\omega_X -\mu -zt-\Delta)\psi_B&=&\frac{g \psi_B/2}{\sqrt{|\psi_A|^2+|\psi_B|^2}}.
\end{eqnarray}
The semiclassical equations have three solutions: (i) $\psi_A=\psi_B=0$ corresponding to MI, (ii) $\psi_A =(g/2)/(\omega_X -\mu -zt) $ and $\psi_B=0$ corresponding to SF(A), and (iii) $\psi_A = 0$ and $\psi_B = (g/2)/(\omega_X -\mu -zt-\Delta) $ corresponding to SF(B). Among the three, SF(B) of (iii) minimizes the energy in Eq.~(\ref{eq:Esemi_lambda}) for $\Delta>0$. Therefore, there is no possibility for the SF(A+B) phase to appear in the semiclassical regime (large $t$ and/or high $\mu$). 

Here we should note that the semiclassical result is consistent with the perturbation theory presented in Sec.~\ref{sec2-2} (small $t$) which gives no possibility of the phase transition from MI to SF(A+B) phase.
However, contrary to the semiclassical equations, the ground state can be SF(A) instead of SF(B), even if $\Delta>0$ for quantum regime with low $\mu$ i.e. weak excitation regime.
The results in the two limits (semiclassical and perturbative) suggest that, at low $\mu$, the ground state will change from MI to SF(A) and eventually to SF(B) as increasing the hopping parameter $t$. In this case, the phase transition from SF(A) with ($\psi_A>0,\psi_B=0$) to SF(B) with ($\psi_A=0,\psi_B>0$) should be the first-order one. This is because, if SF(A) and SF(B) connect smoothly by the second-order phase transitions, other phases, MI or SF(A+B), must exist between SF(A) and SF(B), which seems to be unphysical.

Moreover, another type of the first-order phase transition is possible, which is the transition from the MI0 state (vacuum ground state) to semiclassical SF(B) state. The situation occurs when $E_{\rm semi}(\psi_A=0,\psi_B>0)=E_{{\rm MI}0}=0$ at the minimum of Eq.~(\ref{eq:Esemi_lambda}). Restoring the last term $E_{\rm emit}$ in Eq.~(\ref{eq:Esemi_lambda}), this condition is given by
\begin{eqnarray}
\frac{\omega_X-\mu+\Delta}{2}-\frac{(g/2)^2}{\omega_X-\mu-\Delta-zt}=0, \label{eq:SemiPB_lambda}
\end{eqnarray}
and plotted by a dotted line in Fig.~\ref{fig4} showing a good agreement with the exact numerical solution (thick solid line) in the low-$\mu$ limit.

Physically, the above mathematical arguments can be understood as follows.
Because the photon energy of mode B is less than that of the mode A ($\omega_B=\omega_A -\Delta$), the dressed photon energy $E_{\rm ph}$, namely a sum of the first three terms in Eq.~(\ref{eq:Esemi_lambda}), is reduced for large $n_B$ state if $n_A+n_B$ is fixed. (This is also the reason why $n_B \ge n_A =0$ in all MI phase with $\mathcal{N}_{tot} \ge 1$ and why only SF(B) phase appears in semiclassical regime.)
Reminding that the modes A and B couple, via the dipole transition, to the emitter state $|\alpha=0 \rangle$ and $|\alpha=2 \rangle$, respectively, the large $n_B$ state has a probability $P_2 \equiv \langle \hat{P}_2 \rangle$ larger than $P_0 \equiv \langle \hat{P}_0 \rangle$. 
However, in the case of $P_2 > P_0$, total energy of emitters $E_{\rm emit}$  becomes large, hence cannot be minimized.
Therefore, there can be two competing states minimizing $E_{\rm semi}=E_{\rm ph}+E_{\rm emit}$, which correspond to the two distinct minima: 
(i) one minimizes $E_{\rm ph}$ but maximizes $E_{\rm emit}$ --- which is related to SF(B) phase; 
(ii) the other minimizes $E_{\rm emit}$ but maximizes $E_{\rm ph}$ --- which is related to MI0 and SF(A) phases.
The competition between two states becomes effective only where the total excitation number is small (i.e. $\mu$ is small) such that $\mathcal{N}_{\rm tot} \le \mathcal{O}(1)$. This is due to the fact that $E_{\rm ph}$ and $E_{\rm emit}$ become comparable only for small $\mu$.

\begin{figure}[!tbp] 
\begin{center}
\includegraphics[width=0.4\textwidth]{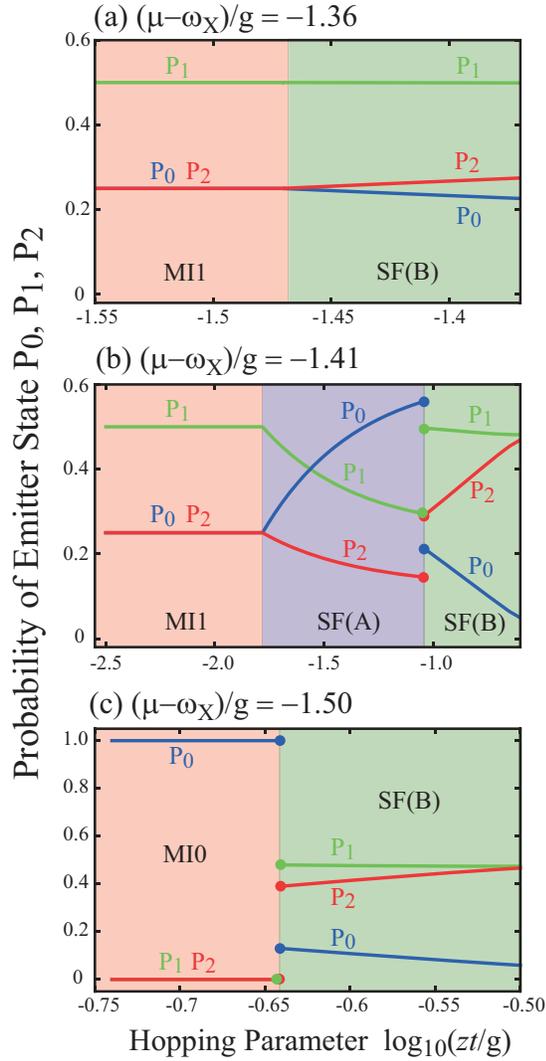} {}
\end{center}
\caption{\label{fig8} (Color online) Probability of emitter states, $P_{\alpha}= \langle \Phi| \hat{P}_\alpha | \Phi \rangle$ with $\hat{P}_\alpha \equiv |\alpha \rangle \langle \alpha |$, are evaluated for the ground state $| \Phi \rangle$ and plotted as a function of hopping parameter for different chemical potential: $(\mu-\omega_X)/g=$ (a) $-1.36$, (b) $-1.41$, and (c) $-1.50$. } 
\end{figure}

Validity of this interpretation is clearly checked in Fig.~\ref{fig8} showing the probability of the emitter states $P_\alpha$ obtained for the ground state as a function of the hopping parameter.
Being consistent with the above physical arguments, we find $P_0>P_2$ in MI0 and SF(A) phases---(i), whereas $P_0<P_2$ in SF(B) phase---(ii).
These two class of states does not connect to each other with a small perturbation, and showing the first-order transition between (i) and (ii).
We should also note here that the MI1, MI2, and the MI states with higher $\mathcal{N}_{\rm tot}$ smoothly connected with SF(B) state via the second-order phase transition. 
This can be understood by seeing $P_{2} \ge P_{0}$ for the MI state $|-,0,n_B \rangle$, since $P_{2}/P_{0}=1+n_B \ge 1$ for $n_B \ge 0$ from Eq.~(\ref{eq:DressdPhoton1}). This means the higher MI states can be in the same class (ii) as SF(B) state.

The origin of the first-order transition in our model is analogous to that of the first-order transition occurs in BH models with two or three component bosons~\cite{Kimura,KimuraJPS,Krutitsky,Kuklov}. In spin-1 BH model~\cite{Kimura}, the first-order transition occurred between two distinct MI and SF states: one is a MI state that minimizes the total spin-spin interaction energy but maximizes the kinetic (hopping) energy, and the other is a SF state that minimize the kinetic (hopping) energy but maximizes the total spin-spin interaction energy.
The first-order transition is possible because, not only the intersite coherence but also the spin state (the internal degree of freedom) show a discontinuous change at the phase transition. 
According to the discussion here, a presence of such competing effect showing distinctive multiple energy minima, due to an additional degree of freedom, seems to be a condition to find a first-order MI-SF transition in general model.

\section{\label{sec3} Test for other emitter models}
Here we discuss the ground states obtained for four different types of emitters (shown in Fig.~\ref{fig9}).
As will be shown here, some models exhibit the first-order phase transition while some does not. Combining the results with those shown in Sec.~\ref{sec2}, we refer to the condition for the first-order MI-SF transition to occur in cavity QED arrays.

\subsection{\label{sec3-1} Four sample models of emitters}
\begin{figure}[!tbp] 
\begin{center}
\includegraphics[width=0.40\textwidth]{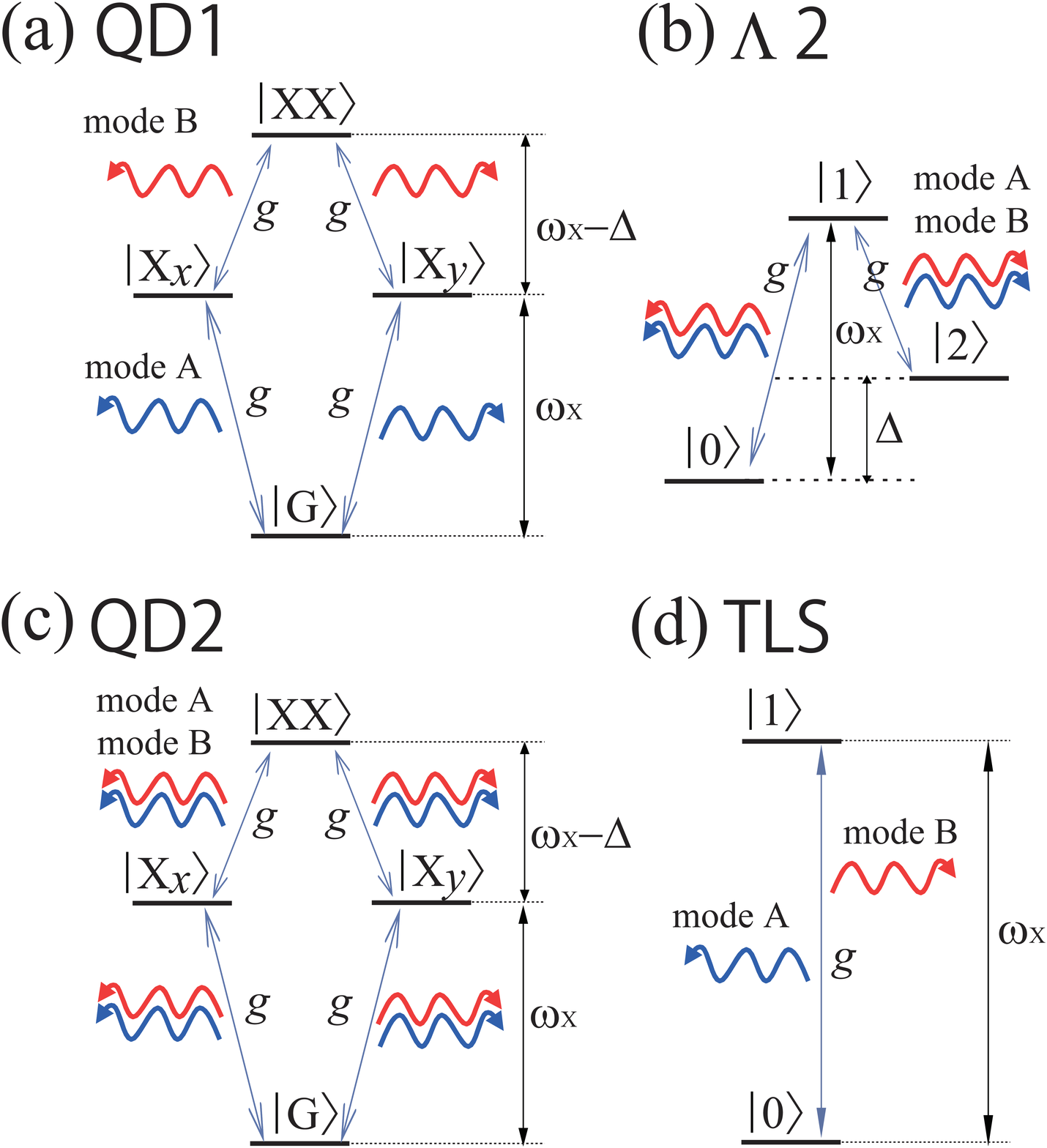} {}  
\end{center}
\caption{\label{fig9} (Color online) Models of three types of emitters considered in Sec.~\ref{sec3}. Here we set frequencies of the cavity modes A ($=\omega_A$) and B ($=\omega_B$) as $\omega_A=\omega_X$ and $\omega_B=\omega_X-\Delta$ in the same way as in Sec.~\ref{sec2}.} 
\end{figure}

Four types of the emitters inside a cavity considered here are shown and labelled by QD1, $\Lambda 2$, QD2, and TLS in Fig.~\ref{fig9}. They are explained as follows.
\par
(QD1) A quantum dot four level system with a biexciton state $| {\rm XX} \rangle$, two exciton states $| {\rm X}_x \rangle$ and $| {\rm X}_y \rangle$, and ground state $| {\rm G} \rangle$, which is coupling with two cavity modes A and B as shown in Fig.~\ref{fig9} (a). Dipole transition is assumed to occur only in case of the resonance; a transition between the biexciton and exciton states is coupled to the mode B ($\omega_B=\omega_X-U$), and that between the exciton and ground states is coupled to the mode A ($\omega_A=\omega_X$). For this model, the Hamiltonian $\hat{h}_{0,i}$ in Eq.~(\ref{eq:GenModel}) is given by 

\begin{eqnarray}
\hat{h}_0&=&\omega_A \hat{a}_A^\dagger \hat{a}_A +\omega_B \hat{a}_B^\dagger \hat{a}_B +(2 \omega_X-\Delta)|XX \rangle \langle  XX |  \nonumber \\
&& +\omega_X\left(|X_x \rangle \langle  X_x |+|X_y \rangle \langle  X_y | \right) -\mu \hat{\mathcal{N}}_{tot}  \nonumber \\
&& +  g \hat{a}_B |XX\rangle (\langle  X_x |+\langle  X_y |)  +{\rm h.c.} \nonumber \\
&& + g \hat{a}_A (|X_x \rangle + |X_y \rangle )\langle  G |    +{\rm h.c.} , \label{eq:Qdot1}
\end{eqnarray}
where 
\begin{eqnarray}
\hat{\mathcal{N}}_{tot} &\equiv& 2|XX \rangle \langle XX|+|X_x \rangle \langle X_x|+|X_y \rangle \langle X_y|\nonumber \\ 
&&+\hat{a}_A^\dagger \hat{a}_A   +\hat{a}_B^\dagger \hat{a}_B. \label{eq:NumQDot}
\end{eqnarray}

($\Lambda$2)
$\Lambda$-type three level emitters are coupling with two cavity modes as shown in Fig.~\ref{fig9} (c), which are different from the one discussed in Sec.~\ref{sec2} (Fig.~\ref{fig2}).
For this model, the Hamiltonian $\hat{h}_{0,i}$ in Eq.~(\ref{eq:GenModel}) is given by replacing the light-matter coupling terms in Eq.~(\ref{eq:HLambda}) by 
\begin{eqnarray}
g (\hat{a}_A+ \hat{a}_B) |1\rangle \langle  0 | +  g (\hat{a}_A+ \hat{a}_B) |1 \rangle \langle  2 |   +{\rm h.c.}. \label{eq:Lambda2}
\end{eqnarray}

(QD2) A quantum dot four level system, which is coupling with two cavity modes A and B with different manners from (QD1), as shown in Fig.~\ref{fig9} (c). 
Both the modes A and B coupled to all the dipole transition irrespective of the resonance condition. 
This model would be more suitable than the model (QD1) for the case where the biexciton binding energy $\Delta$ is of the same order of magnitude as $g$.  For this model, the Hamiltonian $\hat{h}_{0,i}$ in Eq.~(\ref{eq:GenModel}) is given by replacing the light-matter coupling terms (the third and fourth lines in Eq.~(\ref{eq:Qdot1})) by 
\begin{eqnarray}
&& g (\hat{a}_A+ \hat{a}_B) |XX\rangle (\langle  X_x |+\langle  X_y |) +{\rm h.c.} \nonumber \\
&& +  g (\hat{a}_A+ \hat{a}_B) (|X_x \rangle + |X_y \rangle )\langle  G |   +{\rm h.c.}. \label{eq:Qdot2}
\end{eqnarray}

(TLS) A two level atom, which is coupling with two cavity modes A and B through the dipole transition, as shown in Fig.~\ref{fig9} (d).
For this simple model, the Hamiltonian $\hat{h}_{0,i}$ in Eq.~(\ref{eq:GenModel}) is given by 
\begin{eqnarray}
\hat{h}_0&=&\omega_A {\hat a}_A^\dagger {\hat a}_A +\omega_B {\hat a}_B^\dagger {\hat a}_B +\omega_X|1 \rangle \langle  1 |  -\mu \hat{\mathcal{N}}_{tot} \nonumber \\
&& +  g ({\hat a}_A + {\hat a}_B) |1\rangle \langle 0 | +{\rm h.c.}, \label{eq:TLS}
\end{eqnarray}
where 
\begin{eqnarray}
\hat{\mathcal{N}}_{tot} \equiv |1 \rangle \langle 1|+{\hat a}_A^\dagger {\hat a}_A   +{\hat a}_B^\dagger {\hat a}_B. \label{eq:NumTLS}
\end{eqnarray} 

\subsection{\label{sec3-2} Ground state phase diagram ($t \ne 0$)}
\begin{figure*}[!htbp]
\begin{center}
\includegraphics[width=0.7\textwidth]{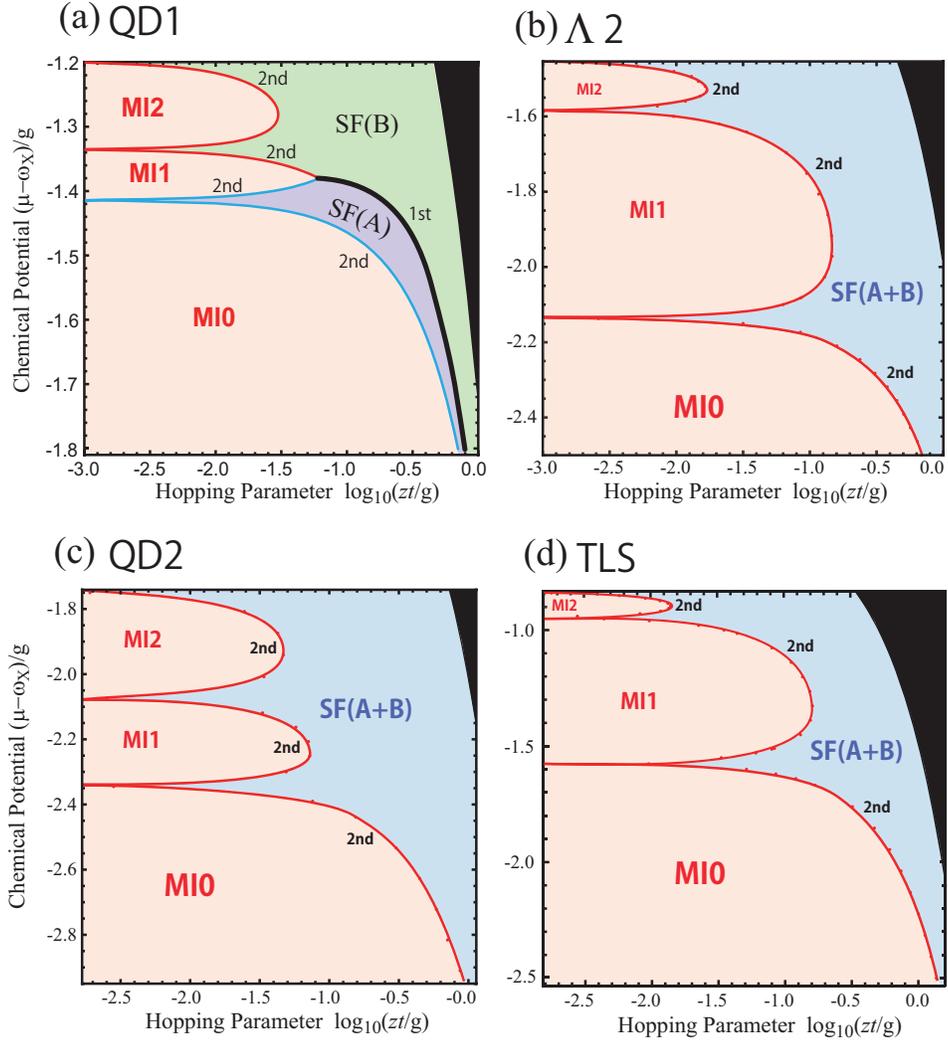}{}
\end{center}
\caption{ \label{fig10} (Color online) Ground state phase diagrams for three different models for emitters: (a) a quantum dot model (QD1) with $\Delta/g=0.75$ and its enlarged view (right panel), (b) a $\Lambda$-type emitters ($\Lambda$2) with $\Delta/g=1.0$, (c) a quantum dot model (QD2) with $\Delta/g=1.0$, and (d) a two-level atom model (TLS) with $\Delta/g=0.5$.
We set a cutoff for photon number so that $n_A,n_B \le n_{\rm max}=16$ for (a), $n_A+n_B \le L_{\rm max} =20$ for (b), $n_A+n_B \le L_{\rm max} =15$ for (c) and (d). The black region is where the number of photons diverges where $\omega_X-\mu-\Delta -zt < 0 $ }
\end{figure*}

In Fig.~\ref{fig10}, we show ground state phase diagrams which are obtained by applying the variational approach in Sec.~\ref{sec2} to the four models, QD1, $\Lambda$2, QD2, and TLS. 
We find the first-order phase transition in QD1 [Fig.~\ref{fig10} (a)], while the phase transition is always of the second order in $\Lambda$2 [Fig.~\ref{fig10} (b)], QD2 [Fig.~\ref{fig10} (c)] and TLS [Fig.~\ref{fig10} (d)]. 
It should be noted that for the latter cases, we found  $\psi_A \ne 0$ and $\psi_B \ne 0$ everywhere in the superfluid phase, which is quite different from the result for the $\Lambda$-type emitters in Sec.~\ref{fig3}. 
On the other hand, the first-order transition from SF(A) to SF(B) is found for QD1 being similar to a model $\Lambda$1, the $\Lambda$-type emitters in Sec.~\ref{fig3}.
According to the result, we conclude that the first-order transition can be found when the different modes are coupling to the different transition processes. 

Here, let's see the origin of the different results between ($\Lambda$1, QD1) and (QD2, $\Lambda$2, TLS) by applying semiclassical arguments to the $\Lambda$2 model being similar to that was shown in the previous section.
In the semiclassical limit $|\psi_A|^2+|\psi_B|^2 \gg1$, the energy of the system is given by
\begin{eqnarray}
E_{\rm semi} &=& (\omega_X -\mu -zt)|\psi_A|^2 +(\omega_X -\mu -zt-\Delta)|\psi_B|^2 \nonumber \\
&&   -\sqrt{2}g |\psi_A + \psi_B| +\mathcal{O}(1),  \ \label{eq:Esemi_lambda2} 
\end{eqnarray}
and the saddle point equations in the semiclassical limit are
\begin{eqnarray}
&& 2 (\omega_X -\mu -zt)\psi_A  -\sqrt{2}g=0, \\
&& 2 (\omega_X -\mu -zt -\Delta)\psi_B -\sqrt{2}g=0,
\end{eqnarray}
where the relative phase between $\psi_A$ and $\psi_B$ is locked to zero for the energy minimization. Therefore, in the semiclassical regime, only the SF(A+B) phase with nonzero $\psi_A$ and nonzero $\psi_B$ is possible, in contrast to the case of ($\Lambda$1, QD1).

The same conclusion is drawn also in the perturbative regime where $|\psi_A|^2+|\psi_B|^2 \ll 1$. If the perturbation theory is applied to the three models (QD2, $\Lambda$2, TLS), one obtains the second-order energy correction to the MI states in a general form
\begin{eqnarray}
&&\delta E_{l}(\psi_A,\psi_B)=zt  |\psi_A|^2 +zt  |\psi_B|^2-z^2 t^2 C_{A,l} |\psi_A|^2 \nonumber \\
&& -z^2 t^2 C_{B,l} |\psi_B|^2 -z^2 t^2 C_{AB,l}(\psi_A \psi_B^\ast+\psi_A^\ast \psi_B)
\label{eq:Emfsecond-2}
\end{eqnarray}
with $C_{AB,l} \ne 0$ [which vanishes for $\Lambda$1 and QD1 as found in Eq.~(\ref{eq:Emfsecond})].
This perturbative expression gives only the possibility that the second-order phase transition from MI to SF states occurs for two modes at the same transition point given by
\begin{eqnarray}
\left|
\begin{array}{cc}
1-zt C_{A,l}  & -zt C_{AB,l} \\
-zt C_{AB,l} & 1-zt C_{B,l} 
\end{array}
\right|=0.
\end{eqnarray}
This indicates that all the Mott phase is surrounded by a SF(A+B) phase.
Therefore, reminding the semiclassical result again, the SF(A+B) phase shows a smooth connection between the perturbative and semiclassical regime, being in contrast to the models ($\Lambda$1, QD1).
According to the considerations, it is reasonable to conclude that there is no first-order phase transition in the three models (QD2, $\Lambda$2, TLS).

\section{\label{sec5} Conclusions}
We investigated the possible quantum phase transition in coupled cavity QED arrays in presence of two cavity modes.
Within the Gutzwiller approximation, which is equivalent to the mean field approximation, the ground state phase diagrams were obtained for types of emitter models with various light-matter interaction configurations, and we found the first-order phase transition between Mott insulator (MI) phase and superfluid phase (or between different superfluid phases) for a certain range of the emitter models.
The first-order phase transition can be detected by the output luminescence spectra and the photon statistics through $g^{(2)}$ measurements~\cite{HBT, Horikiri, Assman}. 
It seems to be necessary, in addition to the multiple components of photons, for the first-order phase transition to appear that the different photon modes should couple with different transitions between emitter states separately.
(A finite difference of the transition energies $\Delta \ne 0$ seems to be important as well.)
In that case, the perturbative regime near the Mott insulating phase and the semiclassical regime can have different types of the ground states, suggesting the first-order phase transition can occur in their intermediate regime.

The origin of the first-order phase transition is analogous to that of spin-1 BH model, where the first-order transition occurs between the states with large difference in the internal spin degree of freedom; MI states includes only the lowest spin state, meanwhile the SF states include the high spin states. Therefore, the kinetic energy and spin-spin interaction energy gives the two competing minima of the total energy in spin-1 BH model.
In case of the two-mode cavity QED arrays, the competition between the dressed photon energy and emitter energy give rise to the first-order phase transition. 
The argument presented here on the origin of the first-order phase transition is useful for future application of these system to possible quantum optical switching devices.

As a future remark, the quantum phase transition of the first order, which is considered here in the thermal equilibrium condition, can occur also in the out-of-equilibrium condition.
We know the thermodynamic phase transition between the different ground states occurs in order to minimize the total free energy of the system of interest. Is there any quantity that can characterize the phase transition even in the out-of-equilibrium condition? In the community of nonequilibrium statistical physics, it has been conjectured that the entropy generation rate of the system of interest should be minimized in the nonequilibrium stationary state, even though this claim has not been clarified yet~\cite{Prigogine}.
Therefore, the entropy generation rate might be such a quantity as to characterize the nonequilibrium phase transition also in the coupled cavity QED arrays.
The problem is interesting also in case of the phase transition of the second order in single-mode cavity QED arrays, since the relation between thermodynamic and non-equilibrium phase transitions has been a recent hot topic.
The related problem is also found in exciton-polariton systems in a semiconductor microcavity both in experiments~\cite{Deng} and theory~\cite{Szymanska, Kamide, Yamaguchi}.
Theoretically, the cavity QED arrays considered here are very useful to investigate this problem, since the density matrix of the system of interest can be fully solved within the mean field approximation irrespective of whether the systems are in the thermal-equilibrium or out-of-equilibrium conditions~\cite{Nissen}.
The application of the cavity QED arrays to that problem will be discussed elsewhere.

\acknowledgements
We thank Y. Kondo, Y. Akutsu, M. Bamba, T. Yuge, R. Nii, T. Ohashi, and K. Asano for the fruitful discussions. We acknowledge the support from KAKENHI (20104008) and the JSPS through its FIRST Program.

\appendix
\section{\label{Sec:Perturbation} Perturbation expansions from Mott insulator phases}
Phase boundaries of the second-order SF-MI transitions are obtained from the perturbation theory from the MI phases.
Considering small perturbation $\hat{\mathcal{V}} \equiv -zt \sum_{i} \left( \psi^\ast_A \hat{a}_{A,i}+  \psi^\ast_B \hat{a}_{B,i} +{\rm h.c.} \right)$, the second-order energy correction for an initial state $l$ (corresponding to one of the MI states) is given by
\begin{eqnarray}
-\sum_{k}\frac{|\langle k|\hat{\mathcal{V}} |l \rangle|^2  }{E_k -E_l },
\end{eqnarray}
where $k$ is an excited state of the unperturbed Hamiltonian (dressed state). Summing up all contributions, we obtained the anaytic expression in Eq.~(\ref{eq:Emfsecond}). The resulting coefficients in Eq.~(\ref{eq:Emfsecond}) for $\Lambda$-type three level emitters are shown in Table~\ref{table1}.

%%%%%%%%%%%%%%%%%%%%%%%%%%%%%%%%%%%%%%%%%%%%%%%%%%%%%%%%%%%%%%%%%%%%%%%%%%%%%%%%%%%%%%%%%%%%      TABLE       %%%%%%%%%%%%%%%%%%%%%%%%%%%%
%%%%%%%%%%%%%%%%%%%%%%%%%%%%%%%%%%%%%%%%%%%%%%%%%%%%%%%%%%%%%%%%%%%%%%%%%%%%%%%
\begin{table*}[thbp]
\caption{\label{table1} Coefficients in the second-order perturbation energy to MI states for $\Lambda$-type three level emitters in Fig.~\ref{fig3}.}
\begin{ruledtabular}
\begin{tabular}{llll}
$l=$ &  unperturbed state  & $C_{A,l}=$ & $C_{B,l}=$  \\
\hline
\hline
%%%%%%%%%%%%%%%%%%%%%%%%%%%%%%%%%%%%%%%%%%%%%%%%%%%%%%%%%%%%%%%%%%%%%%%%%%%%%%
MI0 & $|0,0,0 \rangle$   
 & $\frac{\left(\frac{1}{2} \right)^2}{\omega_{\rm X}-\mu-\sqrt{2}g}+\frac{\left(\sqrt{\frac{1}{2}} \right)^2}{\omega_{\rm X}-\mu}+\frac{\left(\frac{1}{2} \right)^2}{\omega_{\rm X}-\mu+\sqrt{2}g}$ 
& $\frac{(1)^2}{\omega_{\rm X}-\mu-\Delta}$ \\
%%%%%%%%%%%%%%%%%%%%%%%%%%%%%%%%%%%%%%%%%%%%%%%%%%%%%%%%%%%%%%%%%%%%%%%%%%%%%%
\hline
%%%%%%%%%%%%%%%%%%%%%%%%%%%%%%%%%%%%%%%%%%%%%%%%%%%%%%%%%%%%%%%%%%%%%%%%%%%%%%
MI1 & $|-,0,0 \rangle$ 
& $\frac{\left(\frac{1}{2} \right)^2}{-\omega_{\rm X}+\mu+\sqrt{2}g}
+\frac{\left(\frac{1}{2}+\frac{1}{2}\sqrt{\frac{2}{3}}+\frac{1}{2}\sqrt{\frac{1}{6}} \right)^2}{\omega_{\rm X}-\mu-\sqrt{3}g+\sqrt{2}g}$ 
& $\frac{\left(\frac{1}{2} \right)^2}{-\omega_{\rm X}+\mu+\sqrt{2}g +\Delta}
+\frac{\left(\frac{1}{2}+\frac{1}{2}\sqrt{\frac{2}{3}}+\frac{1}{2}\sqrt{\frac{1}{6}} \right)^2}{\omega_{\rm X}-\mu-\sqrt{3}g+\sqrt{2}g-\Delta}$ \\
&& $+\frac{\left(\frac{1}{2}-\frac{1}{2}\sqrt{\frac{2}{3}}-\frac{1}{2}\sqrt{\frac{1}{6}} \right)^2}{\omega_{\rm X}-\mu+\sqrt{3}g+\sqrt{2}g}$
& $+\frac{\left(\frac{1}{2}-\frac{1}{2}\sqrt{\frac{2}{3}}-\frac{1}{2}\sqrt{\frac{1}{6}} \right)^2}{\omega_{\rm X}-\mu+\sqrt{3}g+\sqrt{2}g-\Delta}$\\
%%%%%%%%%%%%%%%%%%%%%%%%%%%%%%%%%%%%%%%%%%%%%%%%%%%%%%%%%%%%%%%%%%%%%%%%%%%%%%
\hline
%%%%%%%%%%%%%%%%%%%%%%%%%%%%%%%%%%%%%%%%%%%%%%%%%%%%%%%%%%%%%%%%%%%%%%%%%%%%%%
MI2 & $|-,0,1 \rangle$  
& $\frac{\left(\sqrt{\frac{1}{6}} \right)^2}{-\omega_{\rm X}+\mu+\sqrt{3}g}
+\frac{\left(\frac{1}{2}+\sqrt{\frac{1}{3}} \right)^2}{\omega_{\rm X}-\mu+\sqrt{3}g-\sqrt{4}g}$ 
& $\frac{\left(\frac{1}{2}+\frac{1}{2}\sqrt{\frac{1}{6}}+\frac{1}{2}\sqrt{\frac{2}{3}} \right)^2}{-\omega_{\rm X}+\mu+\sqrt{3}g-\sqrt{2}g+\Delta}
+\frac{\left(-\frac{1}{2}\sqrt{\frac{1}{3}}+\sqrt{\frac{1}{3}} \right)^2}{-\omega_{\rm X}+\mu+\sqrt{3}g+\Delta}
+\frac{\left(\frac{1}{2}-\frac{1}{2}\sqrt{\frac{1}{6}}-\frac{1}{2}\sqrt{\frac{2}{3}} \right)^2}{-\omega_{\rm X}+\mu+\sqrt{3}g+\sqrt{2}g+\Delta}$ \\
&
&$+\frac{\left(\frac{1}{2}-\sqrt{\frac{1}{3}} \right)^2}{\omega_{\rm X}-\mu+\sqrt{3}g+\sqrt{4}g}$
& $+\frac{\left(\frac{1}{2}\sqrt{2}+\frac{1}{2}\sqrt{\frac{1}{6}}+\frac{1}{2}\sqrt{\frac{3}{2}} \right)^2}{\omega_{\rm X}-\mu+\sqrt{3}g-\sqrt{4}g-\Delta}
+\frac{\left(\frac{1}{2}\sqrt{2}-\frac{1}{2}\sqrt{\frac{1}{6}}-\frac{1}{2}\sqrt{\frac{3}{2}} \right)^2}{\omega_{\rm X}-\mu+\sqrt{3}g+\sqrt{4}g-\Delta}$ \\
\end{tabular}
\end{ruledtabular}
\end{table*}
%%%%%%%%%%%%%%%%%%%%%%%%%%%%%%%%%%%%%%%%%%%%%%%%%%%%%%%%%%%%%%%%%%%%%%%%%%%%%%%%%%%%%%%%%%%%%%%%%%%%%%%%%%%%%%%%%%%%%%%%%%%%%%%%%%%%%%%%%%%%%%%%%%%%%%%%%%%%%%%

\section{\label{Sec:Mollow} Mollow-like side peaks in non-equilibrium and thermal equilibrium situations}
Here, we demonstrate that, when a cavity QED system is in the thermal equilibrium ground state and large coherent field is present, photo luminescence exhibits Mollow-like side peaks only in a low-energy side of the main peak.
This is in contrast to the non-equilibrium case of conventional Mollow triplet where an emitter is driven by a coherent field.
To simplify the discussion, we consider a two-level atom as an emitter.

\begin{figure}[tbp] 
\begin{center} 
\includegraphics[width=0.48\textwidth]{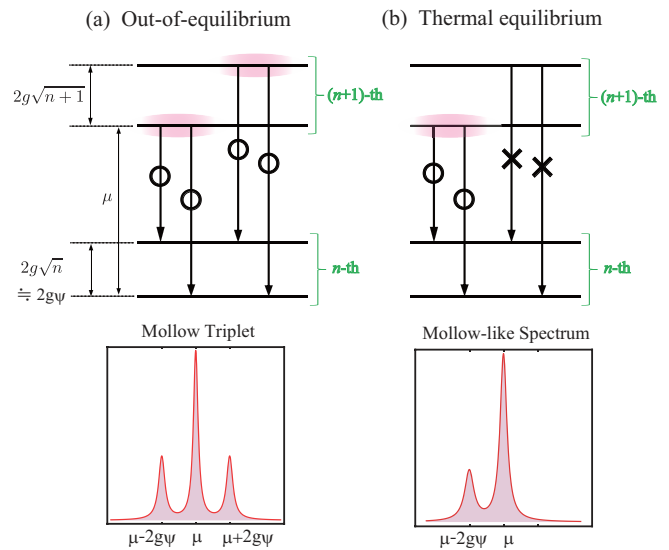} {}  
\vspace{-5mm}
\end{center}
\caption{\label{fig11} (Color online) Schematic explanation for Mollow-like PL spectra the non-equilibrium and thermal equilibrium situation. Here, $(n+1)$-th and $n$-th dressed states are focused for a situation where a two-level system is driven by a coherent field $\psi \sim \sqrt{n} \gg 1 $.}  
\end{figure}

In presence of a coherent field $\psi \sim \sqrt{n} \gg 1$, the energy level of the two-level atom suffers an optical Stark effect and forms a ladder of dressed states.
Consider the $(n+1)$-th and $n$-th dressed states, as in Fig.~\ref{fig11}. 
Such level diagram is often used to account for the Mollow triplet structure of the resonance fluorescence where an atom is driven by a classical field $\psi$. The initial state, before the deexcitation and fluorescence, is found both in the two branches of the dressed states with $(n+1)$ quanta in the driven and out-of-equilibrium system as shown in Fig.~\ref{fig11} (a). This situation gives conventional Mollow triplet spectra (the lower panel). In our case, however, the QED system is driven while it is kept in the thermal equilibrium and occupies the ground state. In this case, the initial state is the unique ground state and found only in the lower branch of the dressed states (so-called lower polariton branch) of $(n+1)$ quanta as shown in Fig.~\ref{fig11} (b). In this case, the resulting PL spectra should display the main peak and only low-energy side peak (the lower panel).

\end{document}